\documentclass[11pt,twoside]{article}
\usepackage{asp2010}
\usepackage{xspace}
\usepackage{epsfig}

\resetcounters

\def\kms{km~s$^{-1}$\xspace}
\newcommand {\arcdeg}   {{\mbox{$^\circ$}}\xspace}

\begin{document}

\title{GASKAP -- A Galactic Spectral Line Survey with the Australian Square Kilometre Array Pathfinder}
\author{John M. Dickey$^1$, S. J. Gibson$^2$, J.F. Gomez$^3$, H. Imai$^4$, P.A. Jones$^5$, N.M. McClure-Griffiths$^6$,
S. Stanimirovic$^7$, J. Th. van Loon$^8$}
\affil{$^1$ School of Maths and Physics, University of Tasmania, Hobart, TAS 7001, Australia}
\affil{$^2$ Physics and Astronomy, Western Kentucky University, Bowling Green, KY 42101-1077, US}
\affil{$^3$ Instituto de Astrofisica de Andalucia, CSIC, Apartado 3004, E-18080 Granada, Spain  }
\affil{$^4$ Graduate School of Science and Engineering, Kagoshima University, 1-21-35 Korimoto, Kagoshima 890-0065, Japan}
\affil{$^5$ School of Physics, University of New South Wales, Sydney, NSW 2052, Australia}
\affil{$^6$ Australia Telescope National Facility, PO Box 76, Epping, NSW 1710, Australia}
\affil{$^7$ Department of Astronomy, University of Wisconsin, 475 N. Charter St., Madison, WI 53706, US}
\affil{$^8$ School of Physical and Geographical Sciences, Keele University, Keele, Staffordshire, ST5 5BG Great Britain}

\begin{abstract}
One of the Survey Science Projects that the Australian Square Kilometre Array Pathfinder (ASKAP) telescope will do in its
first few years of operation is a study of the $\lambda$21-cm line of HI and the $\lambda$18-cm lines of OH in the Galactic Plane and the
Magellanic Clouds and Stream.  The wide-field ASKAP can survey a large area with very high sensitivity
much faster than a conventional telescope because of its focal plane array of receiver elements.  
The brightness sensitivity for the widespread spectral line emission of the interstellar medium depends
on the beam size and the survey speed.  In the GASKAP survey, maps with different resolutions will be synthesized
simultaneously; these will be matched to different scientific applications such as diffuse HI and OH emission, OH masers, and
HI absorption toward background continuum sources.  A great many scientific questions will be answered by
the GASKAP survey results; a central topic is the exchange of matter and energy between the Milky Way disk
and halo. The survey will show how neutral gas at high altitude (z) above the disk, like the
Magellanic Stream, makes its way down through the halo, what changes it experiences along the way,
and how much is left behind.
\end{abstract}
\eject

\centerline{\large \bf Background}
\vspace{.2in}

The Canadian Galactic Plane Survey (CGPS) was a very successful project that had a massive impact on the field of Galactic
radio astronomy.  It was formulated and driven by the staff of the Dominion Radio Astrophysical Observatory (DRAO),
supplemented by an entire generation of young people doing doctoral and post-doctoral research based in Canadian
universities.  The CGPS not only gave astronomers a new and unprecedented view of the Milky Way, it also established
the Canadian scientific community as worldwide leaders in Galactic astronomy.  But the roots of this great success reach
back to a disappointing setback.  More than 20 years ago, the DRAO proposed a new paradigm radio telescope,
called the ``radio Schmidt'' because it was optimized for surveys through a very wide field of view (Landecker and Dewdney,
1991, Dewdney and Landecker 1991).  The proposed telescope was to be made up of a large number (100) of small diameter
(12m) dishes, a concept that is now abbreviated as LNSD.  The LNSD telescope was not funded when it was proposed in the
late 1980's, but it has become the dominant concept for the Square Kilometre Array (SKA).  As a step toward the
SKA, the Australian CSIRO is building a LNSD telescope called the Australian SKA Pathfinder (ASKAP), with 36
dishes of 12m diameter (Johnston et al. 2007).  ASKAP will be an excellent survey instrument in
its own right, because of its innovative multi-receiver system called a focal-plane array (FPA).  
At wavelengths from 18 to 40 cm the FPA gives an instantaneous field of view of 30 square degrees.

The survey science projects (SSP's) that have been selected as the priorities for the first five years of
ASKAP operations are particularly well suited to display the strengths of
the LNSD plus FPA design.  The SSP's are mostly extragalactic continuum and HI line surveys, but
they include Galactic topics such as pulsar timing, transient sources, and Faraday rotation measure surveys.
Carrying on the CGPS model, one of the SSP's is a spectral line survey of the Galactic Plane, Magellanic
Clouds (MC's) and the Magellanic Stream (MS); it goes by the unimaginative but familiar acronym GASKAP, the
Galactic ASKAP survey.  The proposal was written by a consortium of scientists of many nationalities and
many areas of expertise.  Since the proposal was approved, the team of people who will lead the survey has
begun to grow.  Leadership of GASKAP and the other ASKAP surveys is effectively moving to the young generation
of people who did their Ph.D. thesis work on the CGPS and related Galactic H{\tt I} surveys a decade ago.
\vspace{.2in}

\vspace{.1in}
\centerline{\large \bf Survey Science Goals}
\vspace{.05in}
The astrophysical applications of an H {\tt I} and OH spectral line survey of the MW disk and MC's and MS are very
diverse.  Both lines can be seen in thermal emission and absorption, and the OH line in maser emission that
arises in stars at two different stages of evolution.  The simplest attraction of the H {\tt I} emission line is that it traces
the column density of the atomic hydrogen, which is one of the most widespread and abundant constituents of the
interstellar medium (ISM) in the Galaxy.  Thus the 21-cm line is one of the best tracers of Galactic structure and
dynamics, both in the MW disk and in the MC's.  Even when the atomic phase of the gas is a minor component,
as where the hydrogen is mostly ionized or mostly molecular, the 21-cm line may still be the best tracer of the
gas, particularly if molecules like CO are under-abundant or sub-thermally excited, i.e. mostly in the ground
state.  For example, most of our knowledge of the morphology, density, and dynamics of the MS
comes from the 21-cm line, even though most of the surrounding halo gas, and much of the Stream itself, is ionized
(Westmeier and Koribalski 2008, Stanimirovic et al. 2008).

\begin{figure}\label{fig:areas}
\vspace{.5in} \includegraphics[scale=0.5,angle=270]{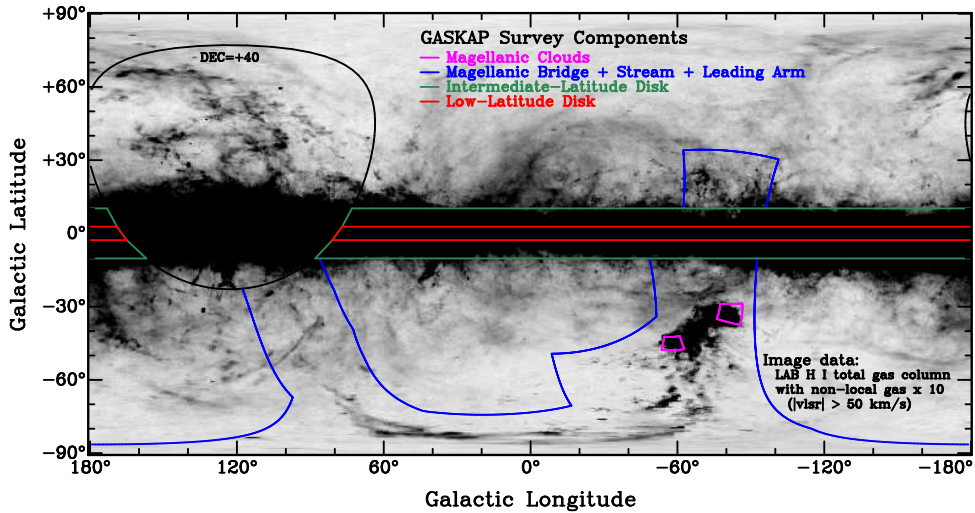}

\caption{The GASKAP survey areas.  The region above $\delta = +40$\arcdeg (upper left) is too far north to be observed
from the ASKAP telescope site in Western Australia.  The thin and thick strips along the Galactic plane are mapped at
different speeds.  The Magellanic Stream area, reaching across the bottom of the figure and through the plane on the right
side, is mapped more rapidly.  The two small areas containing the Magellanic Clouds are mapped in ultra-deep mode, 
given on table 1.}
\end{figure}

In the MC's themselves the atomic hydrogen is the dominant form of the ISM; even more so than in the MW 
where molecular clouds account for a larger fraction of the ISM mass (Hughes et al. 2010).  Comparing 21-cm emission and
absorption spectra shows that the abundance of cool neutral medium (CNM, with kinetic temperature $T_k \sim 50$ K)
relative to warm neutral medium (WNM, with $T_k \sim 6000$ K) is lower in the MC's than in the MW disk (Marx et al. 1997).
This points to a different equilibrium between heating and cooling processes in the gas, probably due to the 
different dust abundance and radiation field intensity as well as the lower abundance of the atoms that dominate
the cooling, C, O, and N.  How the mixture of ISM phases changes with interstellar environment is one of the 
big questions for understanding star formation in different kinds of galaxies, particularly in the early universe
(Wolfire et al. 2003).  

A similar study of the abundance of CNM vs. WNM as a function of galactocentric radius will be possible 
by GASKAP measurements of precision absorption spectra toward background continuum sources
in all directions.  In comparison to the results of the 
CGPS and its cognate surveys the Southern Galactic Plane Survey (McClure-Griffiths et al. 2005) and the
VLA Galactic Plane Survey (Stil et al. 2006), GASKAP will provide more than ten times as many absorption 
spectra at corresponding levels of noise in optical depth (Dickey et al. 2009).  This will be the basis for
a detailed and sensitive longitude-velocity diagram of the cool phase H {\tt I}, in contrast to the l-v diagram of
warm gas that comes from 21-cm emission surveys.

Supplementing the absorption spectra toward continuum background sources, the GASKAP survey will show a remarkable
new picture of H {\tt I} self-absorption (HISA, Gibson et al. 2005), the absorption of Galactic 21-cm emission by 
foreground CNM.  HISA traces preferentially the coolest H {\tt I}, typically with temperatures of 40 K or below, 
where the gas is partly atomic and partly molecular.  Thus mapping HISA shows the transition from CNM to
molecular clouds, and comparison with CO and other molecular tracers plus far-IR dust maps
then clarifies how well each molecule traces the total gas column density in these transition
regions or ``diffuse molecular clouds''.

Moving further into the molecular phase, the diffuse OH emission at $\lambda$18-cm in the Galactic plane will
show the H$_2$ distribution with a new tracer, with different biases from the well studied $^{12}$CO 
and $^{13}$CO emission lines at 115 and 110 GHz.  The OH emission has not been surveyed since the late
1970's (Turner 1979), and GASKAP will improve both the resolution and sensitivity of the emission line maps
by at least a factor of ten over existing data, particularly in the Southern Hemisphere.  The molecular
hydrogen column density corresponding to the OH detection threshold in the GASKAP survey is about 10$^{21}$
cm$^{-2}$, or $A_v \sim 0.6$, which is typical of the conditions where atomic clouds make the transition
to molecular gas.  In addition to 
$\lambda$18-cm emission, the OH lines are often detected in absorption.  The GASKAP survey
will measure optical depths in OH toward the same continuum background sources that provide H {\tt I} absorption
spectra.  Comparison of the two will be a good tracer of the molecular fraction of gas in the clouds, and
of the excitation rate of the OH molecule, that depends strongly on the collision rate.  

OH masers are found in star formation regions and in the outflows from evolved, post-AGB stars.  Both kinds
of masers will be detected in the GASKAP survey with roughly ten times the numbers available from 
all previous surveys combined.  In the LMC, existing surveys have flux limits an order of magnitude
higher than GASKAP will achieve; given the luminosity function this improvement should lead to two orders
of magnitude more maser detections from evolved stars, and many more from star formation regions as well.
Masers detected in the MC's and Magellanic Bridge region will be particularly important for follow-up 
observations with VLBI to study the dynamics of the Magellanic System.

More nearby OH masers will be detected throughout the Galactic disk and particularly in the Galactic
Centre area.  In star formation regions, OH masers are good tracers of outflows associated with the
T Tauri wind from recently formed, massive stars.  The GASKAP results will be complementary to H$_2$O
masers and thermal molecular lines, particularly in the mm-wave band where ALMA will be able to map
the surrounding molecular clouds with excellent sensitivity and resolution.  In post-AGB stars, the OH masers also trace
outflows, but in this case the outflow is associated with the formation of a planetary nebula (PN) and the
transition to a white dwarf star.  Dynamical tracers of this outflow will be crucial for understanding
how the flow is guided to ultimately give the fine scale structure seen in PNe (Zijlstra et al. 2001,
Cohen et al. 2006).
\vspace{.2in}
  
\centerline{\large \bf Survey Parameters}

\begin{figure}\label{fig:baselines}
\hspace{1.5in}\epsfig{file=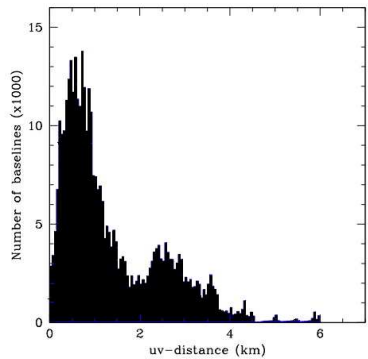,height=2in}
\caption{The distribution of baselines for the ASKAP array.  The two peaks are well suited to the multiple science
goals of a Galactic spectral line survey.}
\end{figure}

The ASKAP telescope has a distribution of baseline lengths that is roughly bimodal, with the largest number in the range
0.3 to 1 km, and a significant second peak between 2 and 3 km (figure 2, taken from Gupta et al. 2008).  The array configuration was
determined primarily by the needs of various extragalactic surveys, but it is excellent for the mixture of Galactic 
science applications of a spectral line survey of the 21-cm and 18-cm lines described in the preceeding section.  The
great advantage of a synthesis survey like GASKAP is
that the baselines can be combined in many different ways using the same data, to obtain many different maps from
a single observation.  By tapering the baselines differently, synthesized beam widths from about 10\arcsec \  to
3\arcmin \ can be generated.  The resulting images have very different brightness sensitivities, with higher sensitivity
corresponding to larger beam size and hence lower resolution.  The GASKAP survey uses this feature of an aperture
synthesis mosaic survey to great advantage, making it effectively four different surveys observed simultaneously.

\begin{center}
{\bf Table 1: GASKAP Survey Area and Speed}
\begin{tabular}{|l|l|r|r|c|}
\hline
 & & {\bf Area} & {\bf Time} & {\bf Speed} \\
{\bf Component Name} & {\bf Location on Sky (Figure 1)} & {\bf deg$^2$} & {\bf hr} & $\!\!${\bf deg$^2$/hr}$\!\!$ \\
\hline
Low Latitude & $|b| < 2.75\arcdeg$, all $\ell$ for $\delta < +$$40\arcdeg$ & 1,496 & 2,493 & 0.60 \\
Intermediate Latitude & $2.75\arcdeg < |b| < 10.25\arcdeg$,  & 4,080 & 1,700 & 2.40 \\
& all $\ell$ for $\delta < +40\arcdeg$ & & & \\
Magellanic Clouds & LMC + SMC deep fields & 94 & 627 & 0.15 \\
Magellanic Bridge & $-135\arcdeg < {\ell_{ms}}^a < +66\arcdeg$, & 5,219 & 2,175 & 2.40 \\
\ \ and Stream &  varying $b_{ms}$$^a$& & & \\
\hline
Total & & 10,779 & 6,995 & \\
\hline
\multicolumn{5}{l}{\footnotesize $^a$Magellanic Stream coordinates (Nidever et al.\ 2008)} \\
\end{tabular}
\end{center}

The survey area will be observed with different speeds, giving three different total integration times per pointing
(dwell time, see table 2) for different regions: 200 hr, 50 hr, and 12.5 hr.  Each of these gives a different
sensitivity, the factors of four in integration time lead to factors of two steps in rms noise.
The brightness temperature sensitivity, which applies to the noise in the emission spectra, is a strong function
of beam size, as shown on table 1.  Thus very faint emission, like the diffuse OH, will be mapped at low
resolution (3\arcmin \ typically), while the much brighter H {\tt I} emission at low latitudes and in the MC's
will be studied with resolution as high as 20\arcsec.  For H {\tt I} absorption toward background continuum sources, the
noise in the optical depth will be set by the fluctuations in the surrounding emission, which must be interpolated
and subtracted from the spectrum toward the source.  For this, the highest possible resolution is needed.
ASKAP will allow resolution as fine as 8\arcsec \ at $\lambda$21-cm, at least in small areas (``postage stamps'').
With this high resolution, emission fluctuations will be very effectively eliminated as a source of 
noise in H {\tt I} absorption spectra.  Figure 3 shows the brightness temperature sensitivity
for the 50 hour dwell time, as a function of beam width.

\begin{figure}\label{fig:sensitivity}
\hspace{1in} \epsfig{file=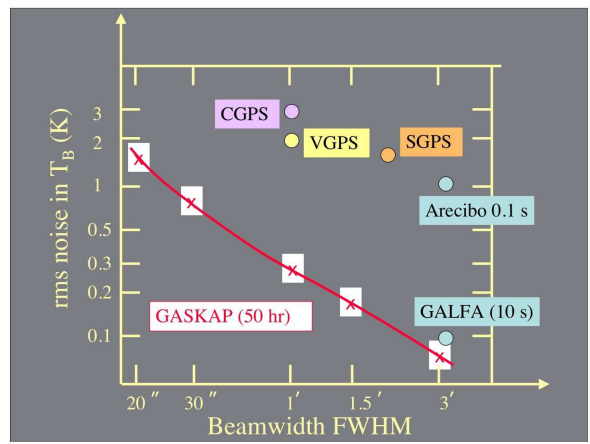,height=2.5in}

\caption{Brightness temperature sensitivity, $T_{rms}$, as a function of resolution (synthesized beam FWHM) for
the GASKAP survey.  This assumes the intermediate dwell time, 50 hr per field, and velocity width of 1 \kms
$\simeq$ 5 kHz.}
\end{figure}

\begin{center} {\bf Sensitivity in Brightness Temperature ($T_{rms}$) and Flux Density ($S_{rms}$)}
\vspace{.1in}

\begin{tabular}{|l|c|c|l|l|l|}
\hline
 &Mapping&Dwell & { $T_{rms}$} [K], $\Delta\nu = 5$~kHz &$S_{rms}$\\
Survey Component & Speed &  Time & ~~~~~for $\theta_{FWHM} =$ &[mJy]\\
 & [deg$^2$/hr] & [hr] &  ~20\arcsec ~~~30\arcsec ~~~60\arcsec ~~~90\arcsec ~180\arcsec & \\
\hline
Magellanic Clouds & 0.15 & 200 & 0.76 ~0.40 ~0.14 ~0.09 ~0.04 & 0.5\\
Low Latitude & 0.60 & 50.0 & 1.51 ~0.81 ~0.28 ~0.19 ~0.08 & 1.0\\
Intermediate Latitude & 2.40 & 12.5 & 3.02 ~1.62 ~0.56 ~0.37 ~0.15 & 2.0\\
Magellanic Stream & & & & \\
\hline
\end{tabular}
\end{center}

The ASKAP spectrometer can provide 16,384 channels over an IF bandwidth of up to 300 MHz, or over several different
narrower bands contained within the total 300 MHz.  By tuning the IF center frequency to approximately 1543
MHz both the 21-cm line and the OH main lines and satellite line at 1612 MHz can be included (but {\bf not} the
1720 MHz OH line).  A good use of the spectral channels is to cover bands of width (in MHz) 7.3 centered
on 1420.4, 4.0 centered on 1612.2, and 6.3 centered on 1666.4.  This gives a Doppler velocity coverage
of $\pm$ 350 \kms and $\pm$ 750 \kms for the OH and H {\tt I} lines, respectively.  The channel spacing is 1.133 kHz in all 
bands, which translates to 0.24 and 0.20 \kms at 1420 and 1667 MHz.  This will give sufficient spectral
resolution to resolve all thermal emission and absorption lines, and all but the narrowest of maser line
components.  As currently planned, GASKAP will use two polarizations, but it will not attempt
to measure all four Stokes parameters, so spectropolarimetry is not an objective of the survey.
\vspace{.2in}

\centerline{\large \bf Data Products}
\vspace{.2in}

The results from the GASKAP survey will be distributed without restrictions as soon as the quality is assured.
This is the policy for all ASKAP survey science projects.  The data products will be spectral line cubes
of the Galactic Plane and of the MC's and MS with various resolutions.  
GASKAP covers most of the Galactic plane (all {\bf except} the region between longitude 80\arcdeg and 160\arcdeg which
is north of $\delta$ = +40\arcdeg) with latitude range $|b| \le 10$\arcdeg.
The specific areas in the Magellanic System that will be covered in the GASKAP survey are given
on table 2.  Source-finding and structure-tracing algorithms will
be applied to the data cubes, particularly for finding weak OH masers and absorption spectra toward faint and
extended continuum sources.  

The observing strategy will seek to provide a uniform sensitivity function over the survey area.
As the FPA technology imposes a strong corrugation on the noise level over the field of view of the individual
antennas, some interleaving of pointings will be inevitable, particularly in the areas observed with the
shortest dwell time.  An approximate goal is to have the noise level change by no more than 10\% over each
section of the survey (table 1). 

An important consideration for the 21-cm emission survey is to fill-in the short uv-spacings,
corresponding to low spatial frequencies, using single-dish data.  The plan is to use
the GASS survey (McClure-Griffiths et al. 2009, Kalberla et al. 2010) for this purpose.
\vspace{.2in}

\centerline{\large \bf References}
\vspace{.2in}

\begin{list}{}{
\setlength{\topsep}{0.0in}
\setlength{\partopsep}{0.0in}
\setlength{\parsep}{0.0in}
\setlength{\itemsep}{0.0in}
\setlength{\leftmargin}{0.2in}
\setlength{\rightmargin}{0.0in}
\setlength{\listparindent}{0.0in}
\setlength{\labelwidth}{0.0in}
\setlength{\labelsep}{0.0in}
\setlength{\itemindent}{-\leftmargin}
}

\item Cohen, R.~J. et al.\ 2006, MNRAS, 367, 541

\item Dewdney, P.E. and Landecker, T.L., 1991, ASPC 19, 415

\item Dickey, J. M.  et al.\ 2009, \apj, 693, 1250.

\item Gibson, S. J. et al.\  2005, \apj, 626, 214

\item Gupta, N., Johnston, S., Feain, I., and Cornwell, T., 2008, ``The Initial Array Configuration for ASKAP'', ATNF internal report.

\item Hughes, A., et al.\ 2010, \mnras, 873 

\item Johnston, S., et al.\ 2007, Publications of the Astronomical Society of Australia, 24, 174 

\item Kalberla, P.~M.~W., et al.\ 2010, arXiv:1007.0686 

\item Landecker, T.L. and Dewdney, P.E. 1991, JRASC 85, 219

\item McClure-Griffiths, N. M. et al.\ 2005, \apjs, 158, 178

\item McClure-Griffiths, N.~M. et al.\ 2009, \apjs, 181, 398

\item Nidever, D.~L., et al. \ 2010, BAAS, 41, 319

\item Stanimirovi{\'c}, S.,  et al.\ 2008, \apj, 680, 276 

\item Stil, J. M. et al.\ 2006, \aj, 132, 1158

\item Turner, B. E. 1979, A\&AS, 37, 1

\item Westmeier, T. \& Koribalski, B.S.\ 2008, \mnras, 388, L29

\item Wolfire, M.G. et al.\ 2003, \apj, 587, 278

\item Zijlstra, A.~A. et al.\  2001, MNRAS, 322, 280

\end{list}
\end{document}